# Fabrication of n$^+$ contact on p-type high pure Ge by cathodic electrodeposition of Li and impedance analysis of n$^+$/p diode at low temperatures

Manoranjan Ghosh,[a)] Pravahan Salunke, Shreyas Pitale, S. G. Singh, G. D. Patra, and Shashwati Sen

*Crystal Technology Section, Technical Physics Division, Bhabha Atomic Research Centre, Trombay, Mumbai 400085, India*

**ABSTRACT**

Fabrication of diode by forming n-type electrical contact on germanium (Ge) and its AC impedance analysis is important for radiation detection in the form of pulses. In this work lithium (Li) metal has been electro-deposited on p-type Ge single crystal from molten lithium nitrate at 260°C. The depth of Li diffusion in Ge was successfully varied by changing the electroplating time as determined by sheet resistance (SR) measurement after successive lapping of Ge surface. Li is found to diffuse up to 500 μm inside Ge by heat treatment of as deposited Li/Ge at 350 °C for 1 hour. A stable n-type electrical contact on Ge with SR ~1Ω/□ and impurity concentration ~3.7x10$^{15}$/cm$^3$ is developed by Li incorporation in p-type Ge crystal showing net carrier concentration ~3.4x10$^{10}$/cm$^3$ and SR ~100 KΩ/□. Acceptor concentration determined from the 1/C$^2$ vs V plot shows similar temperature dependence as found by Hall measurement. The fabricated n$^+$/p junction exhibit ideal diode characteristics with gradual increase in cut off voltage at low temperatures. Under forward bias, junction capacitance mainly comprises of diffusion capacitance (~10μF) showing strong frequency dependence and the impedance is partly resistive resulting in semicircular Cole-Cole plot. Imaginary impedance spectra reveal that the relaxation time for the diffusion of majority carriers decreases at higher temperatures and increased forward voltages. The diode is purely capacitive under reverse bias showing a line parallel to the y-axis in the Cole-Cole plot with frequency independent (100Hz-100MHz) depletion capacitance ~10pF.

[a)]Corresponding author: mghosh@barc.gov.in

# 1 Introduction

The utility of Ge detectors is still expanding despite significant development of alternative detector materials. [1] This is on account of the technological advances in Ge detector and related technologies, which has resulted in new and improved detection concepts. [2,3,4]. The outstanding detection properties of Ge are due to its small band gap as well as very large electron and hole mobilities. [5,6] Additionally, large crystal sizes (> 800 ml) of Ge with the highest purity and a minimum of crystal defects can be grown. [7,8,9] This combined with high energy resolution (~2.5 KeV at 662 KeV) has ensured that Ge as a γ-ray detector material has no alternatives without sacrificing performance in terms of efficiency and energy resolution. [10]

Formation of rugged electrical contacts with low specific resistance on Ge is required for efficient collection of the generated electrical signal in diode detector. [11,12,13] An ideal contact layer will have enhanced electrical conductivity and carrier concentration which allows it to transmit the generated signal efficiently. [14] Planar detectors are usually fabricated by forming electron and hole blocking contacts on opposite faces of intrinsic Ge crystal. Conventionally, Li diffusion is performed to fabricate the $n^+$ contact while boron implantation is used for making a $p^+$ contact. [15,16] Li diffusion is achieved by depositing Li film on Ge surface through thermal evaporation, coating Li containing salt and electro-deposition followed by heat treatment. [17,18,19] Electroplating is usually carried out in a molten mix of salts such as $LiNO_3$ and $KNO_3$. [20] Generally, a eutectic mix is chosen in order to achieve a lower melting temperature (preferably below 300°C). This reduces the possibility of immediate diffusion during plating thereby allowing greater control over the diffusion process. [21] In batteries, Li containing salts such as $LiPF_6$ in ethylene carbonate (EC) and diethyl carbonate (DEC), [22] fluoroethylene carbonate (FEC) along with $KNO_3$ and $LiNO_3$ [23] have been utilized as electrolytes for the electroplating process. $LiNO_3$ in combination with gel polymer was successfully utilized as electrolyte for Li electroplating. [24] However, a mixture of salts and polymer can lead to incorporation of additional impurities in high pure Ge. In this work pure $LiNO_3$ has been employed as an electrolyte to deposit Li on Ge at 260°C.

The electro-deposition approach is found advantageous to obtain stable Li film on large volume and arbitrary shaped Ge. Li/Ge (p-type) contacts thus fabricated should form $n^+$/p junction that exhibit excellent diode characteristics with sub nano-ampere leakage current. Thus it is also important to know the characteristics of the fabricated diode detector. However, detail studies on the properties of electrodeposited Li/Ge contacts are not available in the literature. Further, the Ge diode detector behaves like a capacitor under reverse bias (RB) showing capacitance as low as few pF. Therefore, studying the response of the diode to AC signal is a prerequisite to assess the performance of the diode as gamma detector. In this work, we demonstrate through SR and Hall measurement that electro-deposition of Li on Ge is an efficient route to fabricate rugged $n^+$ contact on Ge with SR ~ 1 $\Omega/\square$. $n^+/p$ junction thus fabricated exhibits excellent diode characteristics having rectification ratio ~ $10^6$. Operating bias, frequencies, capacitance and impedance of the diode for detector application were predicted by temperature dependent capacitance as well as impedance measurement by applying small AC signal.

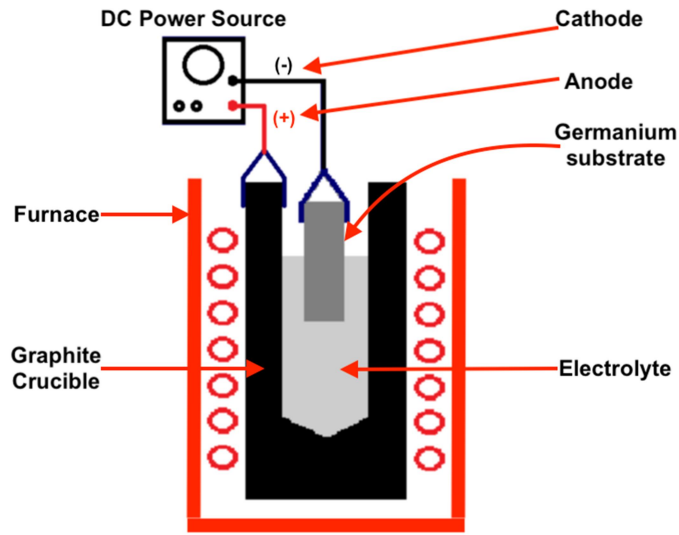

**Fig. 1** (a) Schematic diagram of the Lithium electroplating setup.

## 2 Experimental

A schematic of the electroplating setup is shown in Fig. 1. One graphite crucible of volume 500 ml is used to melt around 100 gm of $LiNO_3$ in a furnace maintained at 280 °C. A flat rectangular piece of a Ge crystal (20x14x2mm$^3$) was immersed in the molten $LiNO_3$ through a holder and stand arrangement. Only half of the length of Ge substrate was immersed in the melt to isolate the metal holder from molten $LiNO_3$. The Ge crystal was maintained as cathode during the electroplating process while graphite crucible acts as anode. A DC power source is used to apply negative voltage on the Ge substrate. In this arrangement $Li^+$ ions approach to the negatively charged Ge cathode and gives rise to high current up to 100 mA due to charge transfer in the process of Li deposition on Ge.

Current-voltage (I-V) characteristics were recorded during electro-deposition to visualize the process (Fig. 2a). Current rises sharply after the application of negative voltage higher than 1V to the Ge cathode and deposition of Li on Ge takes place. A thick Li metal film (~100μm) can be deposited within a very short interval (5 minutes). To study the effect of partial Li diffusion during the deposition, electroplating process was carried out for three different times viz, 2, 5 and 10 minutes. As deposited Li film for the duration of 2 minutes and 10 minutes are shown in Fig. 2b and 2c respectively. After the electroplating process, the coated Ge crystal was washed in methyl alcohol to remove any excess deposition of electrolyte and Li. Gentle lapping of the Ge crystal was performed on an 800 grit SiC abrasive sheet to achieve a smooth flat finished surface by removing the plated layer. Sheet resistance of this surface was measured by 4-point collinear probes connected with Keithley make current source and nano-voltmeter.

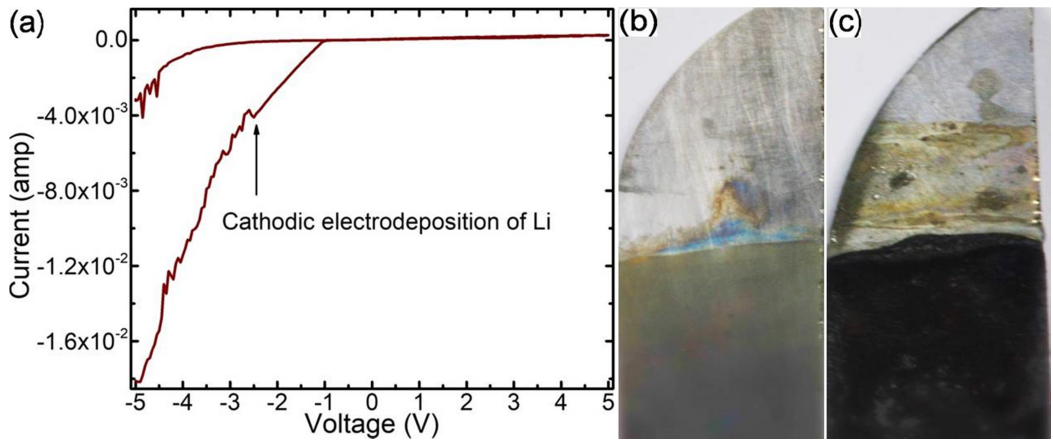

**Fig. 2** (a) Electro-deposition of lithium sharply increases when negative bias of more than 1V is applied on Ge. Li film on the lower part of Ge substrate after 2 minutes (b) and 10 minutes (c) of deposition.

For the fabrication of Li doped Ge ($n^+$)/p-Ge junction diode, a commercially available Ge (100) crystal (acceptor conc. ~$10^{10}$/cc at 77K) from Umicore, Belgium was cut by diamond wheel and ground to a cylinder of diameter 15 mm and length 15 mm. SiC abrasives sheets with grit sizes from 220 to 1500 were successively used for planarization. Lapped surfaces were ultra-sonicated in methanol for 15 min followed by polish etching in 3:1 solution of $HNO_3$:HF (by volume) for 3 min. The cylindrical crystal hold by a metal holder was wrapped by few layers of teflon tape and immersed inside the $LiNO_3$ melt in such a way that only bottom planar surface remains in contact with the melt. After 10 minutes of deposition, the bottom surface of the crystal was electroplated by Li film of thickness around 0.4 mm.

## 3 Results and discussions
### 3.1 Sheet resistance measurement of Li diffused Ge surface along the depth

Diffusion of Li in Ge can be probed by room temperature SR measurement by four collinear probes and using the expression $SR = (V/I)(\pi/ln2)$ where $I$ is the current through two outer probes and V is measured voltage drop between the two inner probes. The diffused Li, supply extra electron to the Ge that participates in electrical conduction thereby reducing the SR. Diffusion of Li takes place during electro-deposition (at 260 °C) without further heat treatment. If the concentration of Li impurities at the surface is infinite, then the depth of diffusion and doping concentration depend on the time of deposition and the constant of diffusion which is a function of temperature. [25] Thus electroplating time can be varied to achieve some degree of control over the depth of diffusion and doping concentration at fixed temperature. For this study, electroplating was carried out for three different durations viz., 2 min, 5 min and 10 min on laminar shape Ge substrates (Fig. 2b) showing SR ~ 100 Ω/□. A thick Li film (0.1-0.4 mm) was deposited on Ge to make the source of diffused impurities infinite. The top flat surfaces were lapped in stages and SR was measured at different depths followed by weight measurements after each successive lapping until the electrical conductivity equals to undoped Ge. As shown in Fig. 3a, the depth of Li diffusion increases with the increase in deposition time at fixed temperature. Li concentration $N(x)$ at depth $x$ can be measured experimentally or deduced from the measured SR values expressed by $(d/dx)(1/SR) = e\mu_{N(x)}N(x)$, where $\mu_{N(x)}$ is the mobility for carrier concentration $N(x)$ and $e$ is the electronic charge. Maximum depth around 250 μm and SR below 4 Ω/□ at the top surface was recorded for 10 minutes of deposition.

For complete diffusion of Li in to Ge lattice, all three samples were further heat treated in air within the same furnace at 350°C for 60 minutes. After heat treatment all three samples exhibit similar depth of diffusion since the Li film thickness was sufficient to consider the source of Li as infinite. A profile of SR vs. depth of 10 minutes electroplated sample after heat treatment has been depicted in Fig. 3b. After successful deposition and heat treatment, Li is

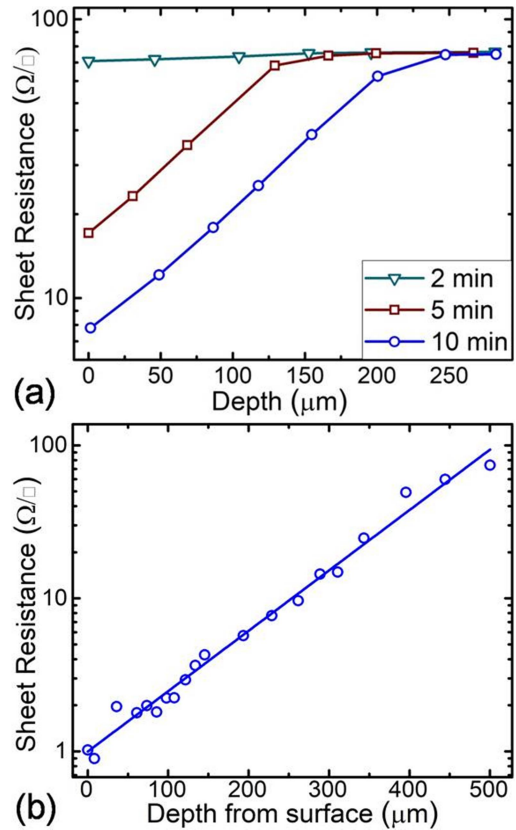

**Fig. 3** (a) Variation of sheet resistance along the depth from the surface of Li deposited Ge crystal for different durations of Li deposition indicated on the graph. b) Sheet resistance along the depth of 10 min Li deposited Ge surface after heat treatment.

found to diffuse up to 500 μm from the surface and SR less than 1 Ω/□ was observed at the top Li doped Ge surface. It increases with the depth from the surface eventually reaching the value shown by pure Ge.

### 3.2 Hall measurement of undoped and Li doped Ge

Basic parameters of a semiconductor such as carrier concentrations, hall coefficient, sheet resistance and carrier mobility of pure and Li doped Ge (heat treated at 350°C for 1 hour) samples were measured by Ecopia make HMS5000 Hall effect measurement system within a range of temperatures between 80K to 300K. A small square piece (10x10x2mm$^3$) with Li doped Ge layer on one side and pure Ge on the other was obtained from the electroplated sample shown

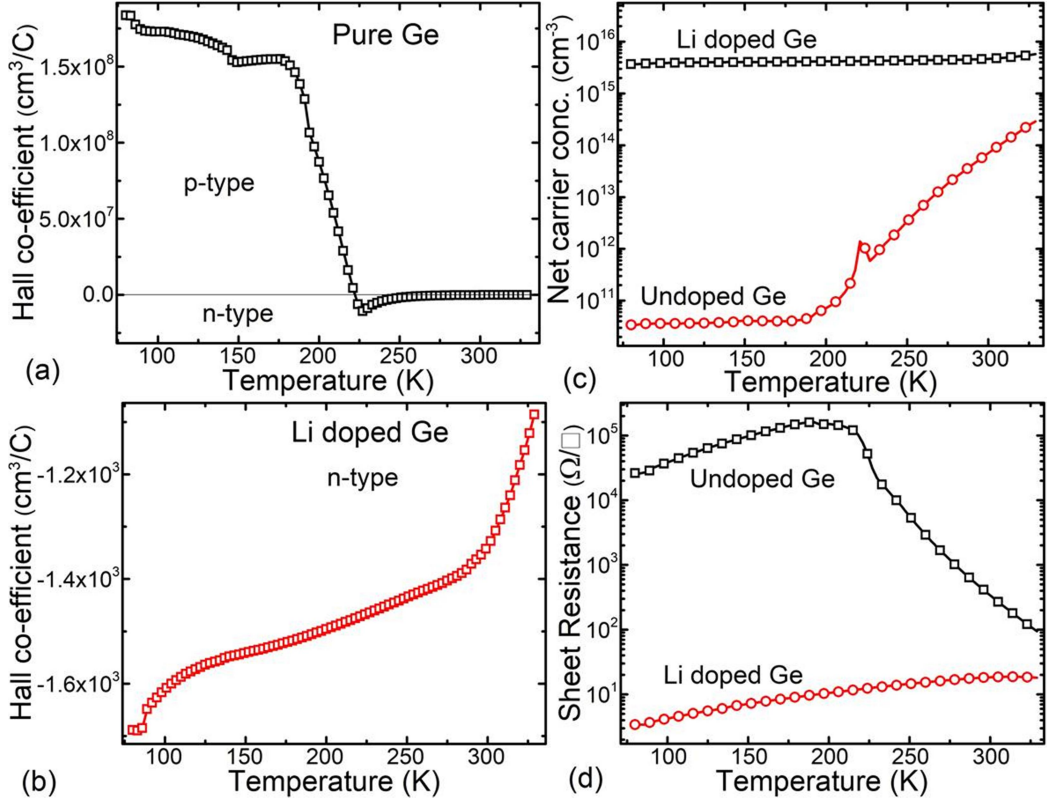

**Fig. 4** (a) The type of conductivity as determined by Hall co-efficient of (a) pure and (b) Li doped Ge. Temperature dependent net carrier concentration (c) and sheet resistance (d) of Li doped Ge layer compared with undoped Ge crystal.

in Fig. 2b for Hall measurements. Surface roughness and strains (such as those caused by lapping) were removed by polish etching according to the procedure described in the IEEE standard. [26] Four contacts were attached on the corners of polish etched square samples at $90^0$ spacing using an Ga-In eutectic in Van der Pauw configuration. [27,28] The Ge crystal chosen for Li electro-deposition shows p-type conductivity in the extrinsic region and exhibit transition to the intrinsic n-type region at temperature ~220K (Fig. 4a). On the other hand Li doped Ge shows n-type conductivity within the temperature range 80-330K as shown in Fig. 4b. The Net carrier concentration of undoped Ge sample was found to be $5 \times 10^{10}/cm^3$ at 80K (Fig. 4c). After Li doping the impurity concentration increases to $5 \times 10^{15}/cm^3$ an enhancement by five orders of magnitude. Also the SR of undoped Ge (>10KΩ/□ at 80K) reduces to ~2Ω/□ after Li doping as seen from Fig. 4d. The hall mobility of undoped and Li doped Ge at 80K shows values 16000

cm$^2$/V.sec and 11000 cm$^2$/V.sec respectively. It gradually reduces as the temperature increases and attains values 1100 cm$^2$/V.sec and 1200 cm$^2$/V.sec at 330K for pure and doped Ge respectively. The p-type undoped Ge crystal displays sharp change in type of conductivity and mobility at 220K above which the electrical conduction is dominated by thermally generated electrons. Whereas no such type transition is observed in case of Li doped Ge because the impurity concentration lies above the intrinsic level.

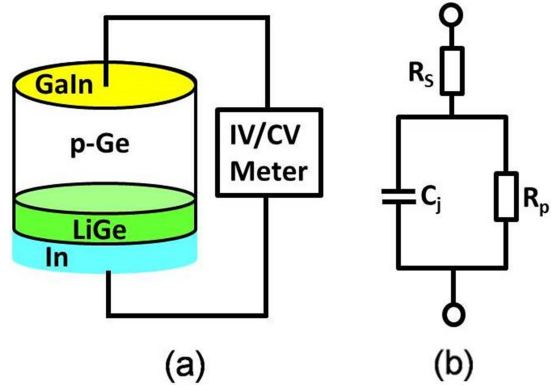

**Fig. 5** (a) Schematic diagram of the Li/Ge n$^+$/p diode (b) Simplified equivalent circuit of the diode.

### 3.3 Current-voltage characteristics of Li doped Ge/Ge diode

As described in experimental section, one planar face of a cylindrical undoped p-Ge crystal of diameter 15 mm and length 15 mm was electroplated with 0.4 mm thick Li (Fig. 5a). For complete diffusion and doping, Li plated Ge crystal was heat treated at 350°C for 1 hr. An n$^+$ electrical contact having Li dopant concentration ~5x10$^{15}$/cm$^3$ (Fig. 4c) and SR ~ 2 Ω/□ has been formed on p-type (impurity level 5x10$^{10}$/cm$^3$) Ge. Li can be incorporated up to 500 μm deep inside the planar Ge surface. Thus an n$^+$/p junction has been formed at the depth where Li concentration equals to the impurity level of p-Ge.

I–V characteristics of the n$^+$/p junction in the temperature range 80–300K shows excellent rectification that indicates formation of high quality junction diode (Fig. 6). The high pure Ge used here exhibits p-type conductivity in the extrinsic region below 220K and the n$^+$/p junction exhibits low reverse current along with exponentially rising forward current, resulting in rectification ratio (RR) as high as 10$^6$. [29] As the temperature increases, both forward and reverse current attains higher values due to thermally generated carriers. [30] In the intrinsic region (above 220K), Ge shows n-type conductivity and the n$^+$/n junction exhibits no rectification. The knee voltage at 80K found to be 0.33V which is a typical value for the n/p Ge diode. It gradually reduces as the temperature increases and exponential rise of the forward current is no more observed after extrinsic (p-type) to intrinsic (n-type) transition of Ge above 220K.

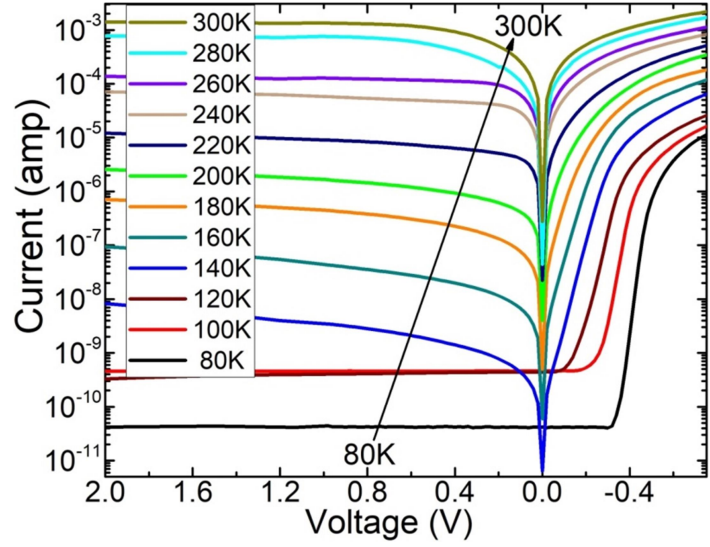

**Fig. 6** I-V characteristics of Li doped n$^+$/p Ge diode within the temperature range 80-300K.

The diode current (I) can be expressed as

$$I = I_o(e^{\frac{qV}{nkT}} - 1) \qquad (1)$$

Where $I_0$ reverse saturation current (A), q electronic charge (1.6x10$^{-19}$C), V applied bias, n is the ideality factor, k is the Boltzmann constant (1.38x10$^{23}$/J/K) and T is the temperature in K. The equation can be directly used to fit the forward current although a mismatch between calculated and the experimental values are possible near the knee voltage.

An alternative method to derive diode parameters were suggested [31] by rewriting the above equation

$$I = I_o e^{\frac{qV}{nkT}}\left(1 - e^{\frac{-qV}{kT}}\right) \qquad (2)$$

The equation (2) can be converted to a linear form

$$ln\left(\frac{I}{\left[1-e^{\frac{-qV}{kT}}\right]}\right) = ln(I_o) + \frac{qV}{nkT} \qquad (3)$$

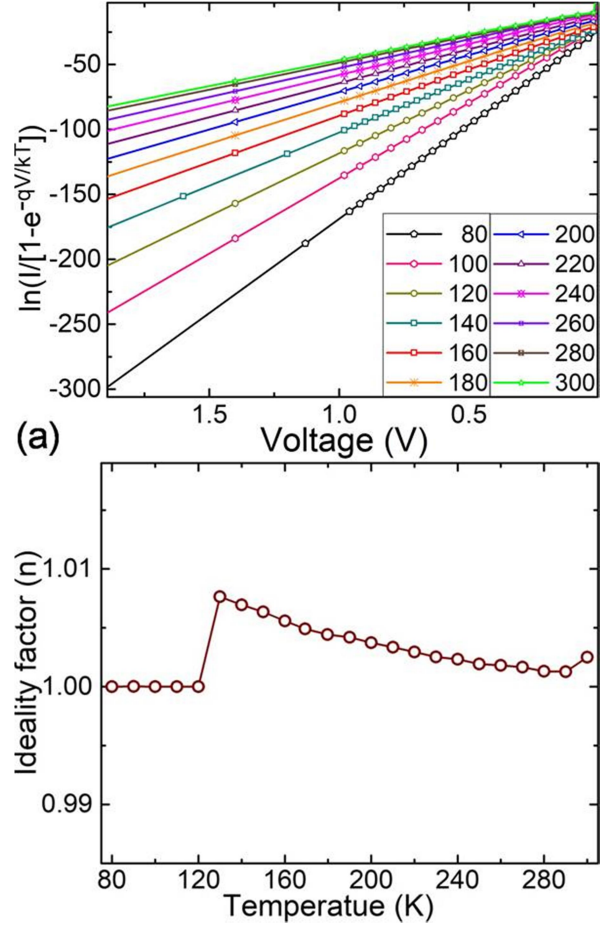

**Fig. 7** (a) $ln(I)$ determined from the experimental I-V results are plotted with the applied bias at various temperatures. (b) Temperature dependent ideality factors determined from the slopes of $ln(I)$ vs $V$ plot.

The measured current $I$ and applied voltage $V$ at various temperatures are used to calculate the left hand side of equation 3 and plotted in Fig. 7a against the reverse voltages. After linear fitting of $ln(I)$ vs $V$ plot, the inverse of the slopes $dV/d[ln\{I/(1-e^{-qV/kT})\}]$ were used to calculate the ideality factor ($n$) for different temperatures. Values of $n$ are found to be sufficiently close to ideal values (1.0 to 1.01) as conformity to the ideal diode behavior (Fig. 7b).

### 3.4 Capacitance-voltage measurements at different temperatures

Li/Ge ($n^+$/p) diode (Fig. 5a) behaves like a capacitor under RB and knowledge of its depletion width and response to AC signal is important for application as detector. [32,33] Capacitance of the p-n junction diode ($C_j$) has been measured at different forward and reverse voltages as well as with frequency at 80K by applying small AC signal of 1 V on top of DC bias voltages.

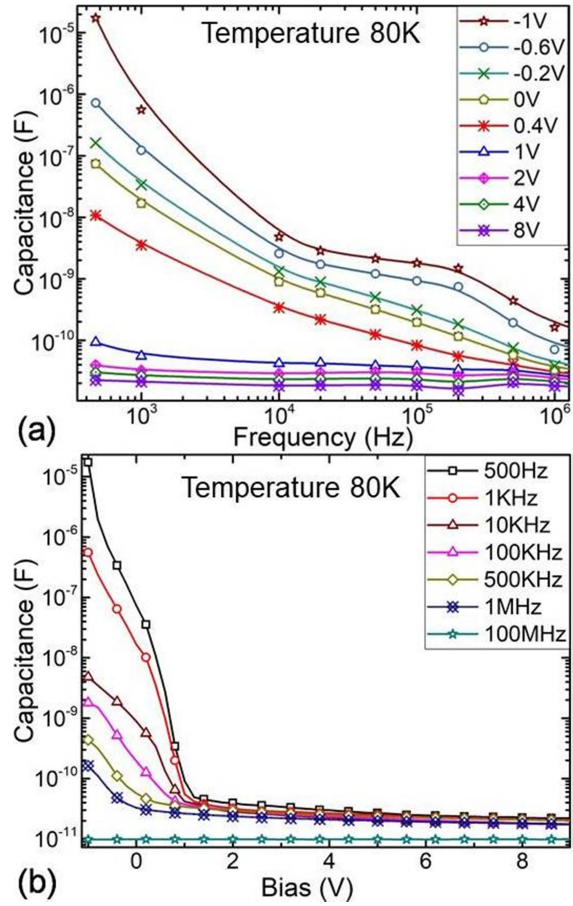

**Fig. 8 (**a) Junction capacitance for different forward and reverse voltages within a range of frequency. (b) Bias voltage dependent capacitance recorded at different frequencies.

As shown in Fig. 8a, higher capacitance values are observed at lower frequencies that gradually decrease with increase in frequency at forward bias (FB). Capacitance value reduces as the forward voltage decreases to zero and becomes few pF under RB condition that has negligible frequency dependence. When plotted against the bias voltages (Fig. 8b), measured capacitance shows higher values (up to 10 µF) under FB condition that converges to very low values (several pF) under RB for all frequencies. This is important because diode detector operates under RB showing low capacitance that remains almost unchanged for a range of frequencies as observed here. The distinct frequency dependence under forward and reverse voltages indicates the presence of two different type of capacitance. The capacitance of the p-n junction ($C_j$) can be understood as a sum of diffusion capacitance ($C_{diff}$) and depletion layer capacitance ($C_{dl}$). At reverse voltage only $C_{dl}$ is present and found to be frequency independent. Whereas, $C_{diff}$ appear to play major role under FB that shows strong frequency dependence. $C_{diff}$ arises from the shift in Fermi level in respect to the band edge due to the injection of carriers into the neutral zones and expressed as [34]

$$C_{diff} = \frac{qI\tau}{2KT}(\tau/2\omega)^{1/2} \qquad (4)$$

where $I=I_0 \exp(qV/2KT)$ is the FB current across the junction and $\tau$ is the average carrier lifetime. The lifetime can be calculated by substituting the experimentally measured FB current and capacitance. $C_{diff}$ is frequency dependent because the injection and rearrangement of carriers does not take place instantaneously. At frequencies above the characteristic frequencies of the recombination process, the carrier densities are no longer able to follow the AC signal and the contribution of $C_{diff}$ to $C_j$ relaxes around 10KHz at FB -1V (Fig. 8a) and also reflected in the Cole-Cole plot (Fig. 10a). $C_{diff}$ reduces further at high frequencies and nearly equals to $C_{dl}$ above 1MHz.

As shown in the I-V characteristics (Fig.4) the forward current at particular voltage of the p-n junction diode increases sharply as the temperature rises that has large impact on the $C_{diff}$ (equation 4). Accordingly the $C_j$ increases at higher temperature as shown in Fig. 9a. At some frequencies (100-200 KHz), $C_j$ decreases at higher FB voltages. It happens when increased charge injection at higher FB enhances the probability of recombination thereby reducing the lifetime ($\tau$).

A depletion layer is formed between two electrical contacts under RB of $n^+/p$ junction due to the sweeping of majority carriers and the diode acts like a parallel plate capacitor. This depletion layer capacitance ($C_{dl}$) has been measured by applying small AC voltage on top of the RB that causes charge accumulation or depletion of the DC space charge. As the $C_{dl}$ originates from free carrier movement, it is almost independent of frequencies and can be written as [35]

$$\frac{1}{C_{dl}^2} = \frac{2(V_{bi} \pm V)}{q\varepsilon_0\varepsilon_s N_A N_D A^2} \qquad (5)$$

where $V_{bi}$ is the built-in voltage, $V$ is the external bias voltage and the signs correspond to the reverse and forward conditions respectively. $\varepsilon_s$ is the dielectric constant of the semiconductor, $\varepsilon_o$ is the dielectric constant of a vacuum ($8.85 \times 10^{-12}$ F/m), $q$ is the electron charge, $A$ the effective area of the diode, and $N_A$ and $N_D$ are the concentration of the acceptors and donors respectively.

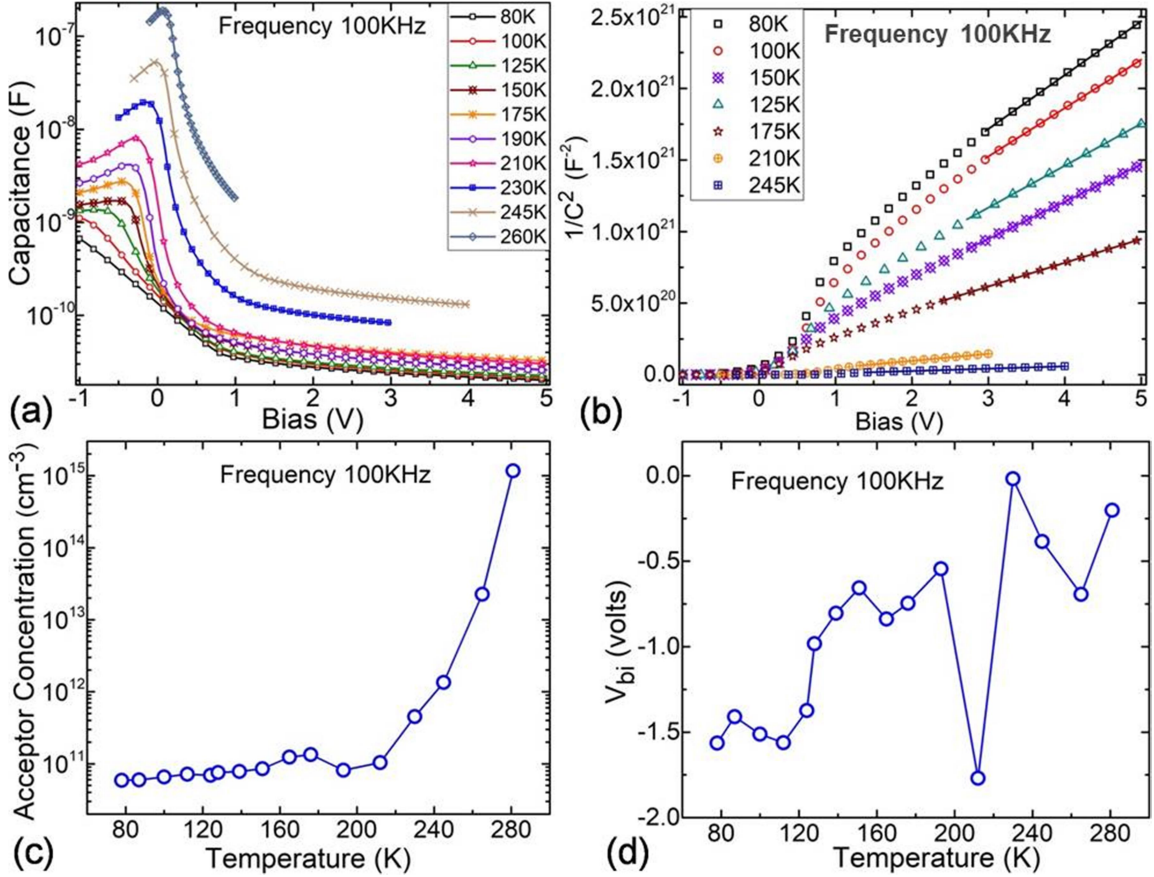

**Fig. 9** (a) Temperature dependence of the junction capacitance for different forward and reverse voltages. (b) $1/C^2$ vs voltage plot at different temperatures shows different slopes after linear fitting. Acceptor concentration and built in voltages ($V_{bi}$) for different temperatures as determined from the $1/C^2$ vs V plot are shown in (c) and (d) respectively.

The plot $1/C^2$ vs V is shown in Fig. 9b. It can be seen that the plot is linear for reverse voltages when the diode is purely capacitive due to the extension of the depletion layers towards the acceptor region (intrinsic crystal) and $C_j$ is equal to $C_{dl}$. The acceptor concentration as well as built in voltages were determined from the slopes and intercepts of linear fitting and plotted in Fig. 9c and 9d respectively. [36] The acceptor concentration as determined from the analysis matches with that determined from the temperature dependent Hall measurement shown in Fig. 4. The built-in potential ($\varphi_{bi}$) is expressed as $(kT/q)ln(N_A N_D/n_i^2)$ and largely depend on $N_A$ and $N_D$. In the present case $N_A N_D$ is less than $n_i^2$ and the built in potential is negative. At higher temperature $N_A N_D$ increases and built in potential assume lesser negative values as shown in Fig.

9d. The capacitance vs voltage measurement was also carried out for different frequencies at 80 K (not shown here). The similar analysis shows that the acceptor concentration does not change with the variation in frequencies.

**3.5 Impedance spectroscopy measurement at different temperatures and bias voltages**

We have seen that $C_j$ of the fabricated $n^+/p$ junction shows relaxation at certain frequencies under FB. The role of $C_{diff}$ and $C_{dl}$ on the relaxation behavior can be clearly understood through the impedance spectroscopy analysis. [37] The complex impedance of the diode along with phase angle was measured over a range of frequencies and the real and imaginary parts of the impedance were extracted. The data can be presented by Cole-Cole plot which is a complex plane representation of real (abscissa) and imaginary (ordinate) part of impedance that normally appears in the form of semi-circles representing the various conduction mechanisms. [38]

An n-p junction can be understood as a parallel combination of resistance ($R_p$) and junction capacitance ($C_j$) comprising of $C_{diff}$ and $C_{dl}$ (Fig. 5b). $R_p$ can be considered as junction resistance in FB and bulk resistance under RB. Apart from that the Ge material offers resistance called dielectric resistance and dielectric capacitance due to contacts. [39] The dielectric resistance is very low compare to the junction resistance especially in case of RB. And the dielectric capacitance due to geometric factor acts on parallel to the diode [40] that plays role at high frequencies. Behavior of low frequency dielectric capacitance could not be determined confidently within the measured range of frequencies. For simplification, the diode is represented only by parallel combination of single resistance ($R_p$) and junction capacitance ($C_j$) placed in series with contact resistance $R_s$ (Fig. 5b).

The impedance of the equivalent circuit is given by [41]

$$Z(\omega) = \left[R_s + \frac{R_p}{1+(\omega R_p C_j)^2}\right] - i\left[\frac{\omega R_p^2 C_j}{1+(\omega R_p C_j)^2}\right] \qquad (6)$$

where $\omega$ is the angular frequency of the ac excitation. By eliminating the angular frequency we obtain the following equation relating the real and imaginary parts of the complex impedance Z, [42]

$$\left[Re(Z) - \left(\frac{R_s+R_p}{2}\right)\right]^2 + Im(Z)^2 = \frac{R_p^2}{4} \qquad (7)$$

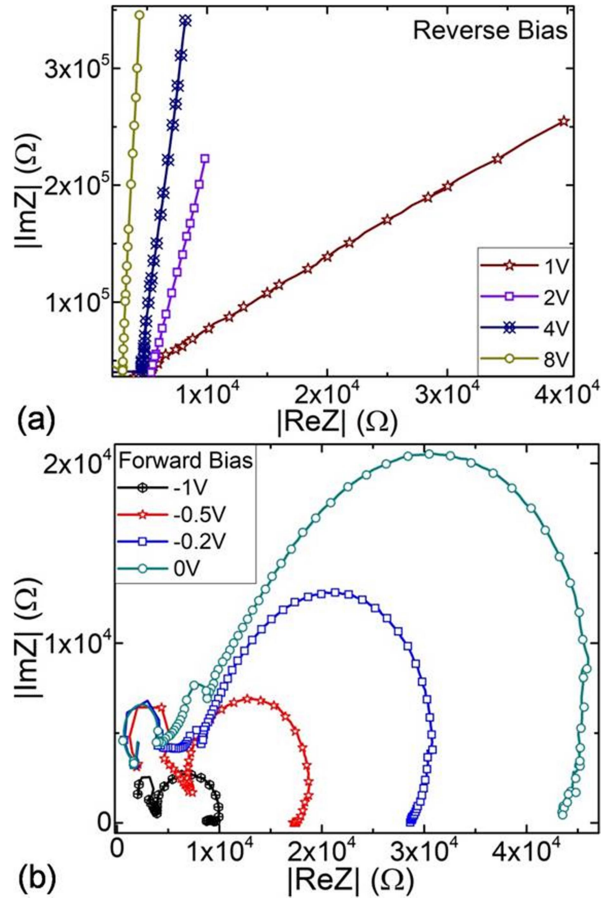

**Fig. 10** Cole-Cole plot of the p-n diode at 80 K for different (a) forward and (b) reverse bias voltages.

The junction resistance $R_p$ is the diameter of the semicircle and can be obtained by interception of semicircle with x-axis. Other parameters along with $R_p$ can be found out by fitting equation 7 with the experimental results.

The Cole–Cole plot of the diode at 80 K for different bias voltages is shown in Fig. 10. In case of RB, the n-p junction behaves like pure capacitor with very high bulk resistance ($R_p$) and the Cole-Cole plot should ideally look like a line parallel to the Y-axis. For low reverse voltages showing relatively lower $R_p$, lines are not fully parallel to Y-axis although no semicircles are observed (Fig. 10a). At reverse voltage around 8 V the diode behaves like pure capacitor and the line is almost parallel to the y-axis. The value of capacitance observed is around few pF due to sweeping of accumulated charge carriers. Only thermally generated minority carries participate in the conduction mechanism that shows high relaxation times.

Mainly two semicircles are seen under FB that indicates the presence of two type of capacitance (Fig. 10b). The small semicircle at high frequency appears due to the dielectric relaxation of Ge. The semicircle due to $C_j$ in the mid-range frequencies plays major role in case of FB and exhibits larger diameter ($R_p$) because of high junction resistance compare to the Ge material itself. Effect of dielectric capacitance at low frequency (below 500 Hz) could not be measured with confidence. The mid-frequency semicircle arises from diffusion of charge carriers that induce change in Fermi level in respect to band edge due to the formation of p-n junction. Junction resistance $R_p$ is minimum for maximum applied FB (-1V) and gradually increases as the bias voltage approaches to zero (Fig. 10b). Every semi-circle corresponding to each bias voltages has its own relaxation frequency ($\omega$) and the relaxation time ($\tau$) given by [43]

$$\tau = \frac{1}{\omega} = R_p C_j \qquad (8)$$

where $\omega = 2\pi v_{max}$, $v_{max}$ is the applied frequency corresponding to arc maximum. The value of junction capacitance can be calculated using the relation $2\pi v_{max} R_p C_j = 1$. [44] Relaxation time increases for higher forward voltages due to increased accumulation as discussed later. The calculated values of $R_p$, $C_j$ and relaxation time ($\tau$) for different temperatures at FB -0.5V are listed in Table 1.

Temperature dependent Cole-Cole plot for different bias voltages are shown in Fig. 11. For better visualization, two different range of temperature are chosen and shown in separate viewgraphs. Under FB voltage of -0.5V, the impedance is found to decrease with increase in temperature which indicates participation of thermally generated electron-hole pair in the electrical conduction. The reduction in bulk resistance is sharper in the temperature range (160-200 K) where carrier concentration quickly increases with the rise in temperature. Also the relaxation times decrease very fast with the increase in temperature in this region which indicates the diffusion of the carriers is a thermally activated phenomenon. Beyond 200 K, Cole-Cole plot is not semicircular due to the absence of n-p junction because the Ge becomes n-type after extrinsic to intrinsic transition. Under RB and very small FB (before the injection become significant) there is formation of depletion layer with high bulk resistance and the diode behaves like a pure capacitor. The bulk resistance depends on the temperature due to generation of thermally generated minority carrier within the depletion region. As the temperature increases, the bulk resistance decreases and the diode deviates from its capacitive nature which is reflected by the semicircular Cole-Cole plot beyond 220 K.

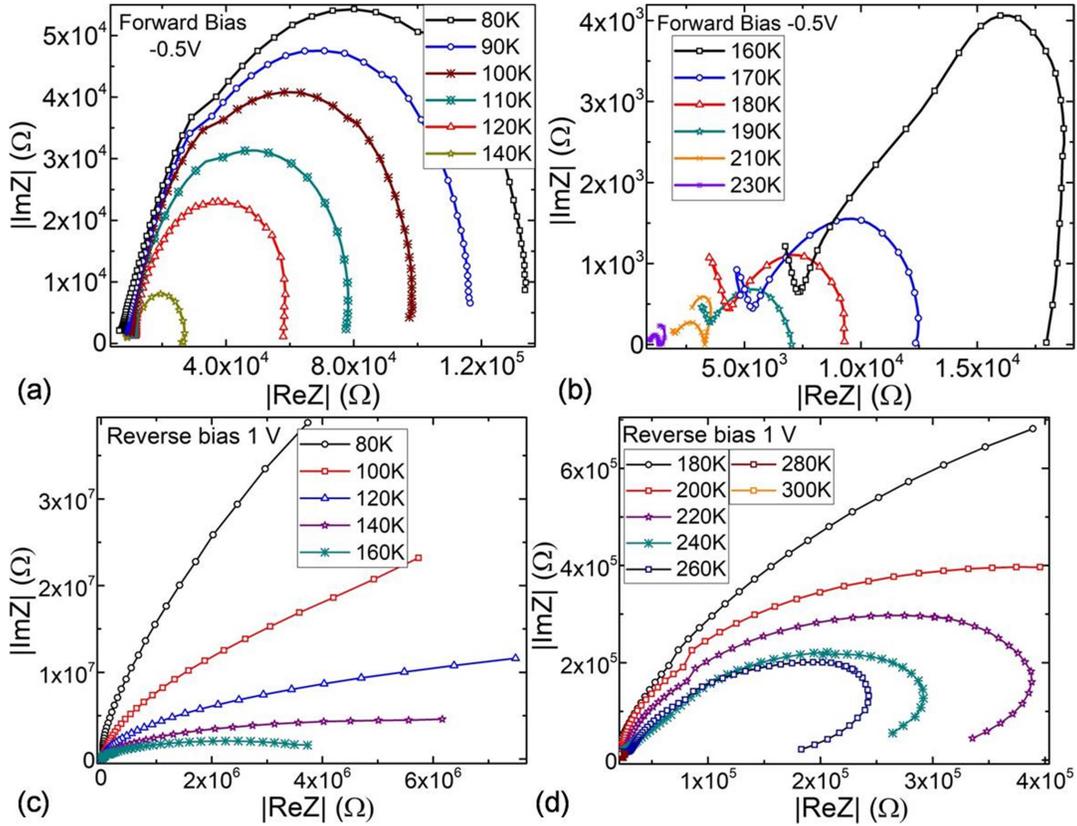

**Fig. 11** Temperature dependent Cole-Cole plot of the p-n diode under forward (a), (b) and reverse (c) and (d) bias voltages. Four separate view graphs are shown to accommodate large change in resistance.

In some cases, imaginary part of impedance is plotted against the frequency to clearly observe the relaxation phenomena in response to AC signal and operational frequency range of the device. Broadly, ImZ decreases for both lower and higher frequency regions with a peak at intermediate frequencies known as relaxation frequency, $f_m$ (Fig. 12). In case of RB at 80 K, there is no movement of charge carriers across the junction and the impedance is purely capacitive. The fluctuating AC signal induces accumulation and sweeping of charge carriers at the electrode which does not show relaxation at lower frequencies. Thus, ImZ vs frequency plot is completely linear and ImZ increases as the frequency decreases (Fig. 12a).

In FB region, majority carrier diffuses across the junction and the impedance is capacitive as well as resistive. The majority carriers contributing to the electrical conduction relax when fails to follow the AC signal at certain frequency and the ImZ decreases at lower frequency in

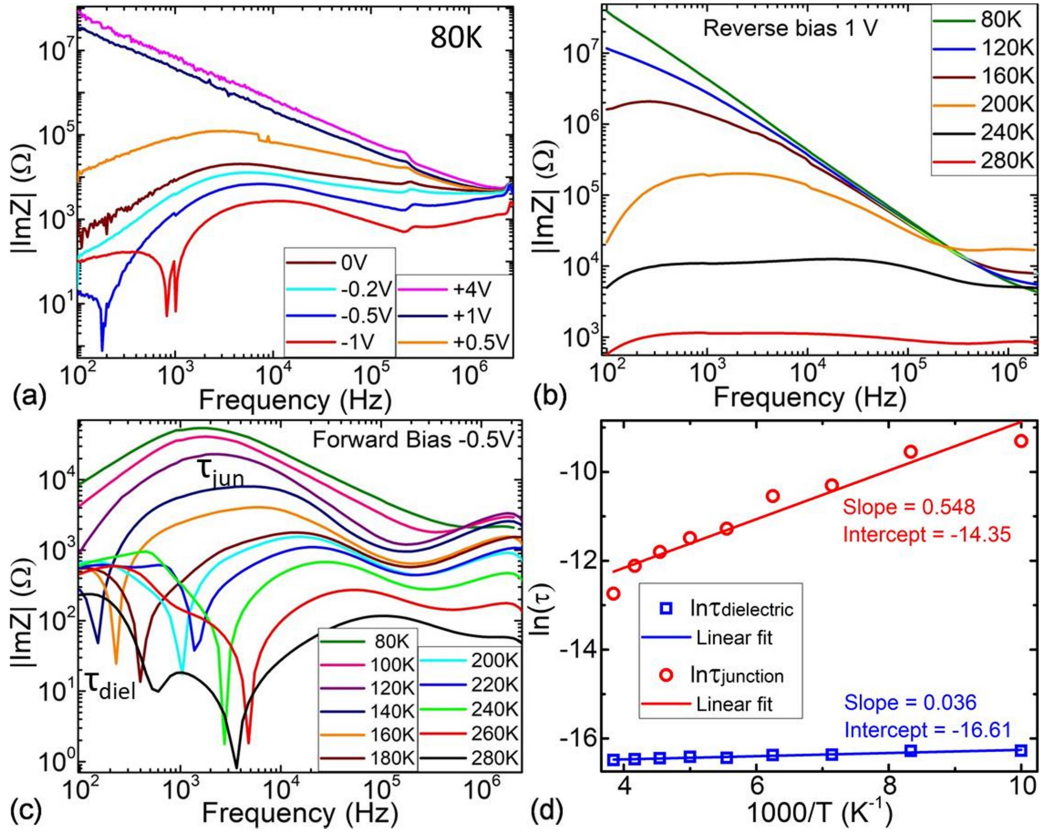

**Fig. 12** (a) Imaginary part of impedance vs. frequency at 80K. The bias voltages are indicated on the graph. The same plot at various temperatures for reverse bias voltage 1 V (b) and forward bias voltage -0.5V (c). (d) The Arrhenius plot of the mid frequency relaxation due to junction capacitance and high frequency dielectric capacitance.

contrary to that observed under RB. The formation of peak indicates that $C_{diff}$ mainly contributes to the $C_j$ and peak position shifts to higher frequencies as the FB voltage increases. In Fig. 12b, ImZ is plotted with frequency at different temperatures for RB 1V. As the temperature rises, thermally generated carries populate in the depletion region that contribute to the electrical conduction producing resistive impedance. And the relaxation of thermally generated carriers occurs at certain frequencies. At very high temperature carrier concentration increases at such level that the impedance becomes purely resistive. In case of FB of -0.5 V (Fig. 12c), $C_{diff}$ has major contribution to the junction capacitance. Impedance vs. frequency spectra exhibits clear peak that is also manifested by semicircular Cole-Cole plot at all temperatures and predict the range of frequencies where AC response to the p-n junction is regular. The high frequency dielectric relaxation is manifested by the hump observed around 1 MHz. This can be attributed to the dielectric capacitance of the material. Peak position for both the process shift to a higher frequency as the temperature increase, which proposes the variation of relaxation time (τ) with temperature and conduction mechanism is mainly due to thermally activated charge carriers [45]. Also, there is sharp dip in the spectrum at lower frequencies. Low frequency dielectric relaxation plays role below this transition frequency and the AC response was found irregular.

**Table 1** Important parameters of the diode under forward bias of -0.5V at different temperatures

| Temperature (K) | $R_p$ (Ω) | $\tau_{diel}$ (sec) | $\tau_{junc}$ (sec) | $C_j$ (F) |
|---|---|---|---|---|
| 80 | 133954 | 9.16x10$^{-8}$ | 9.64x10$^{-5}$ | 3.22x10$^{-10}$ |
| 100 | 97705 | 8.52x10$^{-8}$ | 9.09x10$^{-5}$ | 5.17x10$^{-10}$ |
| 120 | 57880 | 8.48x10$^{-8}$ | 7.13x10$^{-5}$ | 7.39x10$^{-10}$ |
| 140 | 26360 | 7.82x10$^{-8}$ | 3.35x10$^{-5}$ | 1.53x10$^{-9}$ |
| 160 | 17950 | 7.73x10$^{-8}$ | 2.63x10$^{-5}$ | 2.22x10$^{-9}$ |
| 180 | 9300 | 7.28x10$^{-8}$ | 1.26x10$^{-5}$ | 2.69x10$^{-9}$ |
| 200 |  | 7.44x10$^{-8}$ | 1.03x10$^{-5}$ | 3.83x10$^{-9}$ |
| 220 |  | 7.20x10$^{-8}$ | 7.49x10$^{-6}$ | 1.33x10$^{-8}$ |
| 240 |  | 7.06x10$^{-8}$ | 5.49x10$^{-6}$ |  |
| 260 |  | 6.91x10$^{-8}$ | 2.93x10$^{-6}$ |  |
| 300 |  |  | 1.53x10$^{-6}$ |  |

So, there are three types of processes having different response times. The diffusion capacitance observed over longer range of frequencies arises from the concentration gradient of the majority carries and its impedance depends on movement of majority carries across the junction and doping level. Outside this range there is evidence for low and high frequency dielectric relaxation arising from dielectric and geometric capacitance respectively.

The ImZ vs frequency plot shown in Fig. 12c can be used to evaluate the relaxation time τ for both processes related to junction capacitance as well as dielectric capacitance. Peak position (maximum ImZ) in the mid and high frequency region were used to find out the $\tau_{jun}$ and $\tau_{diel}$ respectively using the relation, $\omega\tau = 2\pi f_{max}\tau = 1$. The value of both $\tau_{jun}$ and $\tau_{diel}$ decreases with increasing temperature as shown in Table 1 that indicates temperature dependent electrical relaxation phenomena. Activation energy of these relaxation processes ($\triangle E_\tau$) can be determined using the Arrhenius equation as follows [46],

$$\tau = \tau_0 \exp\left(\frac{\triangle E_\tau}{k_B T}\right) \quad (8)$$

where $\tau_0$ is characteristic relaxation time at infinite temperature and Boltzmann constant $k_B$ = 8.6173303 x $10^{-5}$ eV/K. The plot of ln($\tau$) versus 1000/T for both processes are shown in Fig. 12d. The values of $\triangle E$ for $\tau_{jun}$ and $\tau_{diel}$ are calculated from the slopes of the linear fit and found to be 47 meV and 3 meV respectively.

## 4 Conclusions

Electro-deposition of Li and subsequent heat treatment can be employed to fabricate high quality n-type electrical contact on p-type Ge with low sheet resistance for device application. Some degree of control has been achieved over the depth and concentration of Li diffusion by changing the electroplating time at temperature 260°C. The concentrations of Li incorporation found by Hall measurement agrees well with that determined from the Mott-Schottky analysis of capacitance vs voltage results. The fabricated n-p junction exhibits ideal diode characteristics and temperature dependent analysis of capacitance as well as impedance of the diode at different bias voltages and frequency affirms its applicability as detector. Depletion layer capacitance of few pF observed due to accumulation and sweeping of carries near the electrode at high reverse voltage when no transports of carries observed through the bulk Ge and impedance is completely capacitive. The impedance is partly resistive under FB because of the diffusion of majority carriers across the junction and the developed diffusion capacitance gives rise to semicircular Cole-Cole plot. Imaginary impedance spectra exhibits two peaks originated from junction and high frequency dielectric capacitance and the time of relaxation of both processes increases at lower temperature.

## Acknowledgements

Authors are thankful to Dr. L. M. Pant for his constant encouragement and support. Unconditional help received from all members of Crystal Technology Section is gratefully acknowledged.

## Author contributions

**MG**: Conceptualization, acquisition of data, analysis and/or interpretation of data, writing original drafts, formal analysis and validation, **PS**: Conceptualization, sample preparation, acquisition of data, analysis and/or interpretation of data, writing original drafts, **SP**: conceptualization, developing methodology, review and editing, **SGS**: conceptualization, developing methodology, review and editing, **GDP**: developing methodology, review and editing, **SS**: conceptualization, review, validation and supervision.

## Funding

Open access funding provided by the Department of Atomic Energy, India.

## Declarations

**Competing interests:** The authors have no relevant financial or non-financial interests to disclose.

**Data Availability:** Data will be available on reasonable request to corresponding author.

# References

1. K. Vetter, The Annual Review of Nuclear and Particle Science **57**, 363–400 (2007)

2. M. Yang, Y. Li, Z. Zeng, Y. Tian, T. Xue, M. Zeng, W. Tang, J. Chang, Radiation Detection Technology and Methods **6**, 433–438 (2022)

3. R-M-J. Li, S-K Liu, S-T. Lin, L-T. Yang, Q. Yue, C-H. Fang, H-T. Jia, X. Jiang, Q-Y. Li, Y. Liu, Y-L. Yan, K-K. Zhao, L. Zhang, C-J. Tang, H-Y. Xing, J-J. Zhu, Nuclear Science and Techniques **33**, 57 (2022)

4. M.S. Raut, D. Mei, S. Bhattarai, R. Panth, K. Kooi, H. Mei, G. Wang, Journal of Low Temperature Physics **212**, 138–152 (2023)

5. E.E. Haller, W. L. Hansen, F. S. Goulding, Advances in Physics **30**(1), 93-138 (1981)

6. M. Ghosh. S. Pitale. S. G. Singh, S. Sen, S. C. Gadkari, Bull. Mater. Sci. **42,** 264 (2019)

7. P.C. Palleti, P. Seyidov, A. Gybin, M. Pietsch, U. Juda, A. Fiedler, K. Irmscher, R.R. Sumathi, J Mater Sci: Mater Electron **35**, 57 (2024)

8. W.L. Hansen, E. Haller, MRS Online Proceedings Library **16**, 1–16 (1982)

9. G. Wang, M. Amman, H. Mei, D. Mei, K. Irmscher, Y. Guan, G. Yang, Mater Sci Semicond Process **39**, 54–60 (2015)

10. J. Eberth, J. Simpson, Prog. Part. Nucl. Phys. **60**(2), 283–337 (2008)

11. S. Gaudet, C. Detavernier, A.J. Kellock, P. Desjardins, C. Lavoie, J. Vac. Sci. Technol. **24**, 474–485 (2006)

12. A. Chroneos, D. Skarlatos, C. Tsamis, A. Christofi, D.S. McPhail, R. Hung, Mater. Sci. Semicond. Process **9**, 640–643 (2006)

13. X-H. Meng, G-J. Wang, M-D. Wagner, H. Mei, W-Z. Wei, J. Liu, G. Yang, D-M. Mei, J. Instrum. **14**(2), (2019) P02019

14. M. Ghosh, S. Pitale, S.G. Singh, H. Manasawala, V. Karki, M. Singh, K. Singh, G.D. Patra, S. Sen, Materials Science in Semiconductor Processing **121**, 105350 (2021)

15. R.H. Pehl, R.C. Cordi, IEEE Trans. Nucl. Sci. **22**(1), 177-177 (1975)

16. S. Uppal, A.F.W. Willoughby, J.M. Bonar, N.E.B. Cowern, T. Grasby, R.J.H. Morris, M.G. Dowsett, J. Appl. Phys. **96**, 1376–1380 (2004)

17. R. H. Pehl, R. C. Cordi, F. S. Goulding, IEEE Transactions on Nuclear Science **19**(1), 265-269 (1972)

18. H.J. Fiedler, L.B. Hughes, T.J. Kennett, W.V. Prestwich, B.J. Wall, Nuclear Instruments and Methods **40**(2), 229-234 (1966)

19. R.D. Baertsch, R.N. Hall, IEEE Trans. Nucl. Sci. **17**(3), 235–240 (1970)

20. W. L. Hansen and E. E. Haller, IEEE Transactions on Nuclear Science **28**(1), 541-543 (1981)

21. R. Robertson, T.J. Kennett, Nuclear Instruments and Methods **98**(3), 599-600 (1972)

22. B. Thirumalraj, T.T. Hagos, C-J. Huang, M.A. Teshager, J-H. Cheng, W-N. Su, B-J. Hwang, Am Chem Soc **141**, 18612−18623 (2019)

23. Y. Shuai, Z. Zhang, K. Chen, J. Loua Y. Wang, Chem. Commun **55**, 2376-2379 (2019)

24. M. Liu, Z. Cheng, K. Qian, T. Verhallen, C. Wang, M. Wagemaker, Chem Mater **31**, 4564−4574 (2019)

25. B. Pratt and F. Friedman, Journal of Applied Physics **37**, 1893 (1966)

26. IEEE Standard Test Procedures for High-Purity Germanium Crystals for Radiation Detectors, IEEE Std. 1160-1993, INSPEC Accession Number: 4471071, Nuclear Instruments and Detectors Committee of the IEEE Nuclear and Plasma Sciences Society 1993

27. G. Yang, K. Kooi, G. Wang, H. Mei, Y. Li, Appl. Phys. A **124**, 381 (2018)

28. G Yang, D. Mei, J. Govani, G. Wang, M. Khizar, Appl. Phys. A **113**, 207–213 (2013)

29. A.A. Alarabi, O. Çiçek, H. Makara, et al., J Mater Sci: Mater Electron **35**, 957 (2024)

30. V.V.N. Obreja and A.C. Obreja, Phys. Status Solidi A **207**(5), 1252–1256 (2010)

31. S. Pitale, M. Ghosh, S.G. Singh, H. Manasawala, G.D. Patra, S. Sen, Materials Science in Semiconductor Processing **13**0, 105820 (2021)

32. E. E. Haller, W. L. Hansen and F. S. Goulding, IEEE Transactions on Nuclear Science **19**(3), 271-274 (1972)

33. D. Dai, M. J. W. Rodwell, J. E. Bowers, Y. Kang, M. Morse, IEEE Journal of Selected Topics in Quantum Electronics **16**(5), 1328-1336 (2010)

34. M. L. Lucia, J. L. Hernandez-Rojas, C. Leon, I. Martil, Capacitance measurements of p-n junctions: depletion layer and diffusion capacitance contributions Eur. J. Phys. **14,** 86 (1993)

35. A. Straub, R. Gebs, H. Habenicht, S. Trunk, R. A. Bardos, A. B. Sproul, A. G. Aberle, J. Appl. Phys. **97**, 083703 (2005)

36. G.R. Neupane , M. Bamidele, V. Yeddu, Journal of Materials Research **37**, 1357–1372 (2022)

37. A. K. Jonscher, Solid-State Electronics **36**(8), 1121-1128(1993)

38. J. Bisquert, G. Garcia-Belmonte, Electronics Letters **33**(10), 900-901(1997) IET Digital Library, https://digital-library.theiet.org/content/journals/10.1049/el_19970565

39. S.K. Ghosh, K. Mallick, J Mater Sci: Mater Electron **34**, 1804 (2023)

40. J. Bisquert, L. Bertoluzzi, I. Mora-Sero, G. Garcia-Belmonte, The Journal of Physical Chemistry C **118**(33), 18983-18991 (2014)

41. N. Kumar, S. Chand, Physica B: Condensed Matter **599**, 412547 (2020)

42. A. Rihani, N. Boutabba, L. Hassine1, S. Romdhane, H. Bouchriha, Synthetic Metals **145**, 129–134 (2004)

43. S.S. Fouad, G.B. Sakr, I.S. Yahia, D.M. Abdel-Basset, F. Yakuphanoglu, Physica B **415**, 82–91 (2013)

44. S. Yuvaraj, V.D. Nithya, K. S. Fathima, C. Sanjeeviraja, G.K. Selvan, S. Arumugam, R. K. Selvan, Materials Research Bulletin **48**(3), 1110-1116 (2013)

45. A. Ray, P. Maji, A. Roy, S. Saha, P. Sadhukhan, S. Das, 2019 Mater. Res. Express **6**, 1250h4 (2019)

46. I.S. Yahia, M. Fadel, G.B. Sakr, S.S. Shenouda, F. Yakuphanoglu, W.A. Farooq, Acta Physica Polonica A **120**(3), 563 (2011)